# Spectral Characterization of a Microbolometer Focal Plane Array at Terahertz Frequencies

D. Jang[1], M. Kimbrue[1], Y. J. Yoo[1,2], and K. Y. Kim[1]

[1]*Institute for Research in Electronics and Applied Physics, University of Maryland, College Park, MD 20742*

[2]*Thorlabs Imaging Systems, Sterling, Virginia 20166, USA*

**Abstract**

We have developed a method to characterize the spectral response of an uncooled microbolometer focal plane array at a broad range of terahertz (THz) frequencies (4–50 THz). This is achieved by using a spectrum-tailored blackbody radiator as a broadband THz source and measuring its spectral power with a Fourier transform infrared (FTIR) interferometer. With an additional measurement with a pyroelectric detector as a reference, the spectral response of the microbolometer relative to the pyroelectric reference is obtained with a signal-to-noise ratio of 100 over a >50 THz bandwidth.

Terahertz (THz) frequency radiation at 0.3–30 THz holds great promise for applications in biochemical spectroscopy, biomedical imaging, surveillance, and industrial inspection [1-3]. These applications greatly benefit from real-time THz imaging especially with focal plane arrays (FPAs) [4-6]. Among many types of FPAs, uncooled microbolometer FPAs have been favorably used for video-rate THz imaging [7, 8]. Microbolometers are originally designed for thermal imaging at the long wavelength infrared (LWIR) region at 7–14 μm, but they retain sensitivity at THz frequencies and have been effectively used for THz beam profiling and imaging at longer wavelengths (14–300 μm) [9-13]. These standard IR microbolometers provide noise equivalent power (NEP), the incident power that yields a signal-to-noise ratio (SNR) of unit, in a range of 200–300 pW/$\sqrt{Hz}$ at 2.5–4.3 THz frequencies [4, 7, 10, 14], which is significantly higher than their typical level of ~pW/$\sqrt{Hz}$ at the LWIR band [9, 10].

Microbolometers more sensitive at THz frequencies have been lately developed by NEC [14, 15], INO [16, 17], and CEA Leti [18, 19]. In those microbolometers, THz absorption is enhanced by optimizing the absorber thickness and/or resonant cavity gap [15, 17, 19], tuning the sheet resistance of absorbers [15, 17], or using antenna-coupled cavities [17, 19]. These THz-targeted microbolometers provide improved THz sensitivity with NEP of 25−90 W/$\sqrt{Hz}$ at 2.5–4.3 THz [17, 19].

Both standard IR and THz microbolometers have been mostly characterized at specific THz frequencies, using far infrared gas lasers pumped by $CO_2$ lasers [9], quantum cascade lasers (QCLs) [10, 11], and femtosecond-laser-based THz sources [18]. Individual bandpass THz filters have been also used with broadband THz sources to determine the spectral response at discrete THz frequencies [13]. Fourier transform infrared (FTIR) spectrometers have been also used to spectrally characterize microbolometers [19-21]. However, this method is mostly limited to examine individual absorbing/transmitting films, substrates, or micro-cavity structures, not to characterize the spectral response of fully operating microbolometers.

In this paper, we present a simple method to determine the overall spectral response of a functioning microbolometer at broad THz frequencies. For this, we use a FTIR spectrometer with a spectrum-tailored blackbody emitter as a broadband THz source and a pyroelectric sensor as a reference detector. The pyroelectric detector has a nearly flat spectral response over a wide range of wavelengths and thus can provide a reference spectrum.

In this experiment, we examine a commercial vanadium oxide ($VO_x$) uncooled microbolometer camera (FLIR, Tau 2 336). This microbolometer FPA was previously used to characterize broadband THz radiation emitted from two-color laser produced plasmas [12, 13]. This microbolometer camera can provide THz images with 336 × 256 pixels (17 μm pitch) and stream 14-bit images at a rate of 60 frames/sec with a camera link expansion board (FLIR) and a frame grabber (NI, PCIe-1433). The noise equivalent differential temperature (NEDT) of the microbolometer is specified as <50 mK at 7.5–13.5 μm, but its sensitivity at 0.1–20 THz is largely unknown.

Our experimental setup to characterize the microbolometer is shown in Fig. 1. A lab-built Michelson-type FTIR spectrometer is used to characterize the microbolometer array. A silicon carbide (Globar™) blackbody radiator is used as a broadband THz emitter. It has an active area of 3 mm × 4.4 mm with emissivity of 0.8, capable of reaching ~1700 K with 20 W electric power. The radiation is quasi-collimated by a parabolic reflector, and its power and divergence angle are controlled by an iris diaphragm. A 90-μm-thickness black polyethylene (BPE) filter is used to tailor the radiation spectrum toward the THz region. The spectrally filtered beam is then split into two by an undoped silicon (Si) (>10 kΩ·cm) beamsplitter of ~280 μm thickness. After reflected by gold mirrors, the two beams are recombined and focused by a 90° off-axis parabolic (OAP) mirror onto either pyroelectric or microbolometer detector.

First, the radiation spectrum is characterized with a lithium tantalite ($LiTaO_3$) pyroelectric detector (Spectrum Detector Inc., SPI-A-62-THZ). It has a sensing area of 2 mm × 2 mm, with NEP of 0.4 nW/$\sqrt{Hz}$ at 10 μm. This pyroelectric sensor has an organic black coating, which has a relatively flat response curve over a broad range of optical and infrared frequencies. The radiation beam from the SiC source is optically chopped at 10 Hz, and the signal from the pyroelectric detector is fed into a lock-in amplifier for detection. The pyroelectric detector is calibrated at optical wavelengths (532 nm, 650 nm, and 800 nm), which all provides a similar responsivity of $70 \times 10^4$ V/ W at a chopping frequency of 13 Hz. In addition, the pyroelectric detector shows a linear response over a wide range of peak-to-peak voltages (≤3.5 V). Care was taken not to saturate the pyroelectric signal.

Figure 2(a) shows a sample FTIR interferogram obtained with the pyroelectric detector, averaged from 10 repetitive scans. The mirror in one of the FTIR arms is translated with a step size of 0.3 μm, yielding a temporal step of 2 fs over a time window of 4 ps. The baseline lock-in voltage is measured to be 165 mV, and this corresponds to input beam power of ~50 μW. We note that multiple interference regions were observed due to reflection from both surfaces of the Si beamsplitter, and we chose the strongest interferogram obtained from the central region.

The resultant Fourier-transformed spectrum is shown in Fig. 2(b). The temporal step of 2 fs yields a ~250 THz detectable bandwidth with a spectral resolution of ~0.25 THz. The radiation spectrum reaches up to ~100 THz with peak radiation at 15 THz as shown in Fig. 2(b). This large THz bandwidth is obtained with a proper combination of radiation intensity and BPE filter thickness. BPE can effectively cut down high frequency THz radiation as its transmission drops nearly exponentially with increasing frequency.

Figure 3 shows a simulated beam spectrum obtained from 1700 K blackbody radiation filtered by a 90-µm-thickness BPE window with additional absorption by a Si substrate and water vapor in the air. Here the transmission curves of the BPE and Si are obtained from combined FTIR measurements with commercial (Shimadzu, IR Prestige21) and lab-built spectrometers operating at >12 THz and <12 THz frequency ranges, respectively. The simulated spectrum reasonably well reproduces the measured one in Fig. 2(b). The spectral dips at 21 THz and 43 THz observed in both Figs. 2(b) and 3 are attributed to strong THz absorption in BPE due to C–H bending vibrations. Note that the periodic oscillations observed in the side wings of the interferogram in Fig. 2(a) is caused by BPE absorption mostly at 21 THz. The absorption dip at 18 THz is due to Si but not clearly observed in Fig. 2(b). In addition, many oscillations observed at 22−50 THz in Figs. 2(b) and 3 are due to absorption in the Si and BPE windows, as well as by water vapor in the air. In particular, the double absorption dips at 46 and 49 THz and the noise-like fast oscillations at <15 THz in Figs. 2(b) and 3 are attributed to THz absorption by water vapor.

Once the reference spectrum is obtained, the pyroelectric detector is replaced by the microbolometer while all the other experimental conditions remain the same. No additional filter is added or removed. Figure 4 shows typical focused THz beams obtained with three different time delays between the two beams in the interferometer—(i) >1 ps (two beams separated far away), (ii) 24 fs (two beams destructively interfering), and (iii) 0 fs (two beams constructively interfering). The insets show their background-subtracted images.

To obtain a FTIR interferogram, the beam intensities within a 2 mm × 2 mm region of interest (ROI) around the beam center are selected and spatially integrated so that both the pyroelectric and microbolometer sensing areas are matched. Figures 5(a) shows the integrated signal plotted as a function of the relative delay in the interferometer, obtained with the same time step (2 fs) and scan range (4 ps) as before. The microbolometer images shown in Fig. 4 correspond to the temporal delays marked with (i), (ii), and (iii) in Fig 5(a). The resulting Fourier transformed power spectrum is shown in Fig. 5(b).

Finally, the spectral response curve of our microbolometer is obtained by dividing the measured microbolometer spectrum by that of the pyroelectric detector. The resulting spectral response is shown in Fig. 6.

As expected, our standard IR microbolometer exhibits a strong response at the LWIR regime (20–40 THz). Beyond 50 THz, it yields little or no response. However, it still retains sensitivity at 5–20 THz although the response drops with decreasing frequency. The sensitivity reaches its lowest around 5 THz and then increases with decreasing frequency at <5 THz. Note that our pyroelectric detector has a non-flat spectral response at <10 THz, which is caused by reduced THz absorption by the organic coating. This artifact is corrected by applying a calibrated spectral curve of our pyroelectric detector at <10 THz. The response curve is cut-off below 4 THz and above 65 THz due to poor SNRs. Nonetheless, the Tau 2 microbolometer is still sensitive at <5 THz (see Fig. 5(b)) although it is ~100 times less sensitive compared to its peak at 21–25 THz. This result is consistent with the previous measurements with far infrared laser and QCL sources [9, 10].

There is a spectral dip observed at 40 THz in the response curve. This dip is also shown in Fig. 5(b), and it is not related to BPE absorption at 43 THz. We speculate that the 40 THz dip corresponds to one edge (~7.5 μm) of the microbolometer's detection wavelength, and it may result from a quarter-wave optical cavity design commonly adopted for efficient absorption at the LWIR [21]. In addition, a weak dip at 18.3 THz is caused by THz absorption in the Si protective cover window of Tau 2. This Si window is also partially responsible for the reduced sensitivity at 10–20 THz. We also believe there are unknown coatings and materials that further limit the detection range, especially at ~5 THz where the camera exhibits the lowest sensitivities.

In this experiment, the spectral resolution is 0.25 THz, mainly limited by the scanning range (4 ps). Although it is sufficient to characterize the overall spectral response of the microcomputer, a longer scanning range can be utilized to provide a more accurate response curve. In addition, nitrogen purging can be adopted to minimize THz absorption by water vapor at <20 THz and 40-55 THz.

In conclusion, we have characterized the spectral response of the commercial room-temperature microbolometer camera (Tau 2, FLIR) at 4-50 THz frequencies. The microbolometer is not designed for THz applications, but it shows a relatively good sensitivity at 8–20 THz. The microbolometer, however, yields little or no response at ~5 THz. At <5 THz, the microbolometer becomes responsive again, but ~100 times less sensitive compared to its peak response at 21–25 THz. In this experiment, the microbolometer is tested rather over a broad range of THz frequencies reaching 100 THz. If the response at much lower THz frequencies (for instance <20 THz) is desired, then one can tailor the blackbody radiation spectrum toward lower THz frequencies with a thicker BPE filter and brighter beam intensity. Indeed, our characterization method is flexible and can be easily adopted to inspect all kinds of THz detectors and FPAs at broad THz frequencies.

This work was supported by the National Science Foundations under Award No. 1351455 and the Air Force Office of Scientific Research (AFOSR) under Award No. FA9550-16-1-0163.

**Figure Captions**

**Figure 1.** Schematic of lab-built FTIR Michelson interferometer consisting of a SiC blackbody radiator, an iris diaphragm, an optical chopper, a black polyethylene (BPE) filter, a high-resistivity silicon beamsplitter (BS), gold flat mirrors, an off-axis parabolic (OAP) mirror, and two switchable detectors (microbolometer and pyroelectric). Additional instruments including a chopper controller, a lock-in amplifier, a motional control system, a frame grabber, and a data adequation computer are not shown.

**Figure 2.** (a) FTIR interferogram of a spectrum-tailored blackbody emitter measured with a pyroelectric detector averaged from 10 scans. (b) Fourier transformed spectrum on a log scale (linear in the inset plot). The dips at ~21 THz and ~43 THz are due to strong THz absorption in the black polyethylene (BPE) filter.

**Figure 3.** Simulated THz spectrum (black solid line) obtained from the product of Planck's blackbody radiation at 1700 K (black dash line) and the transmission curves of a black polyethylene (red dash-dot line) of 90 μm thickness, a silicon substrate (green dash-dot-dot line), and atmospheric air with moderate water vapor (gray dot line).

**Figure 4.** Two collinear broadband THz beams focused onto the microbolometer focal plane array with the two beams (i) separated far away in time (>1 ps), (ii) interfering destructively with a relative delay of ~24 fs, and (iii) interfering constructively with 0 fs delay. The insets show the images with their background subtracted.

**Figure 5.** (a) FTIR interferogram obtained with the microbolometer (FLIR, Tau 2 336). The images in Fig. 4 are obtained at relative time delays of (i) >1 ps, (ii) 24 fs, and (iii) 0 fs. (b) Resultant Fourier transformed spectrum on a log scale (linear in the inset plot).

**Figure 6.** Measured spectral response of Tau 2 (FLIR) microbolometer camera on a log scale (linear in the inset plot) obtained from the ratio of the response curves in Fig. 5(b) to Fig. 2(b), with a spectral calibration applied for the pyroelectric signal at <10 THz.

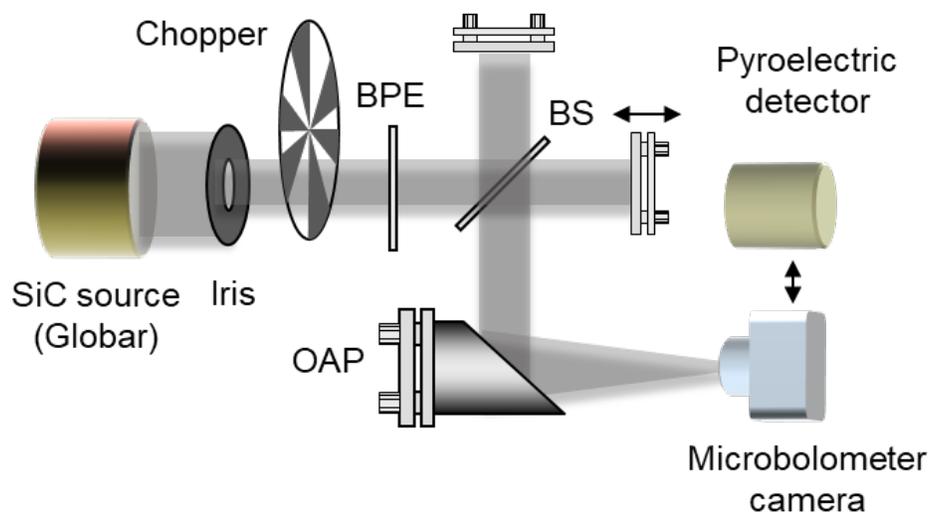

Fig. 1

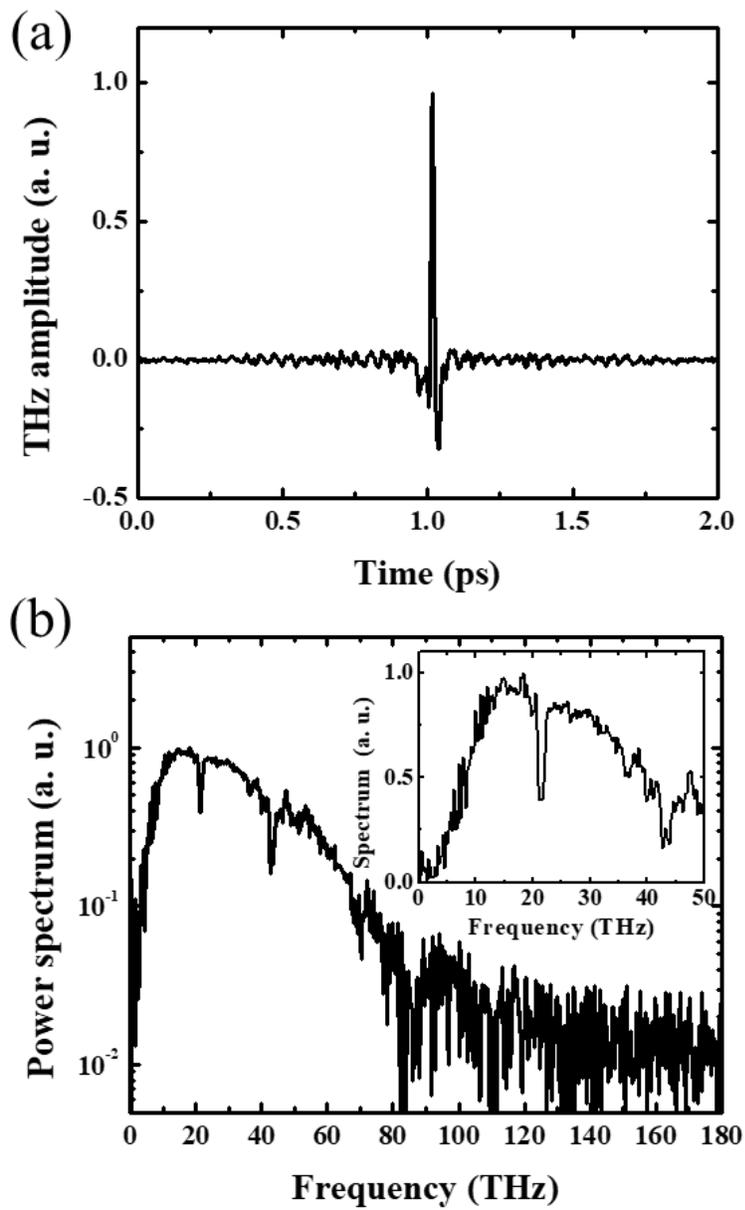

Fig. 2

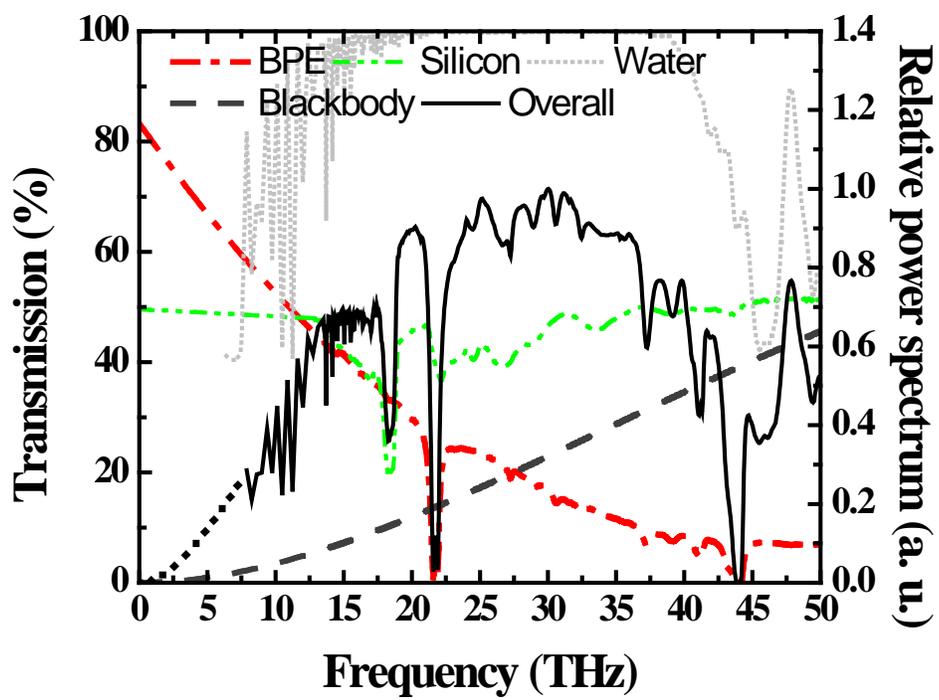

Fig. 3

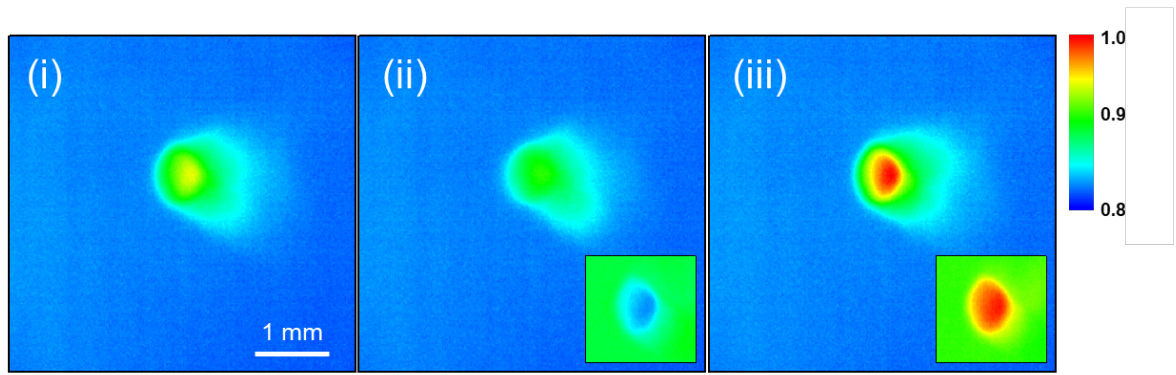

Fig. 4

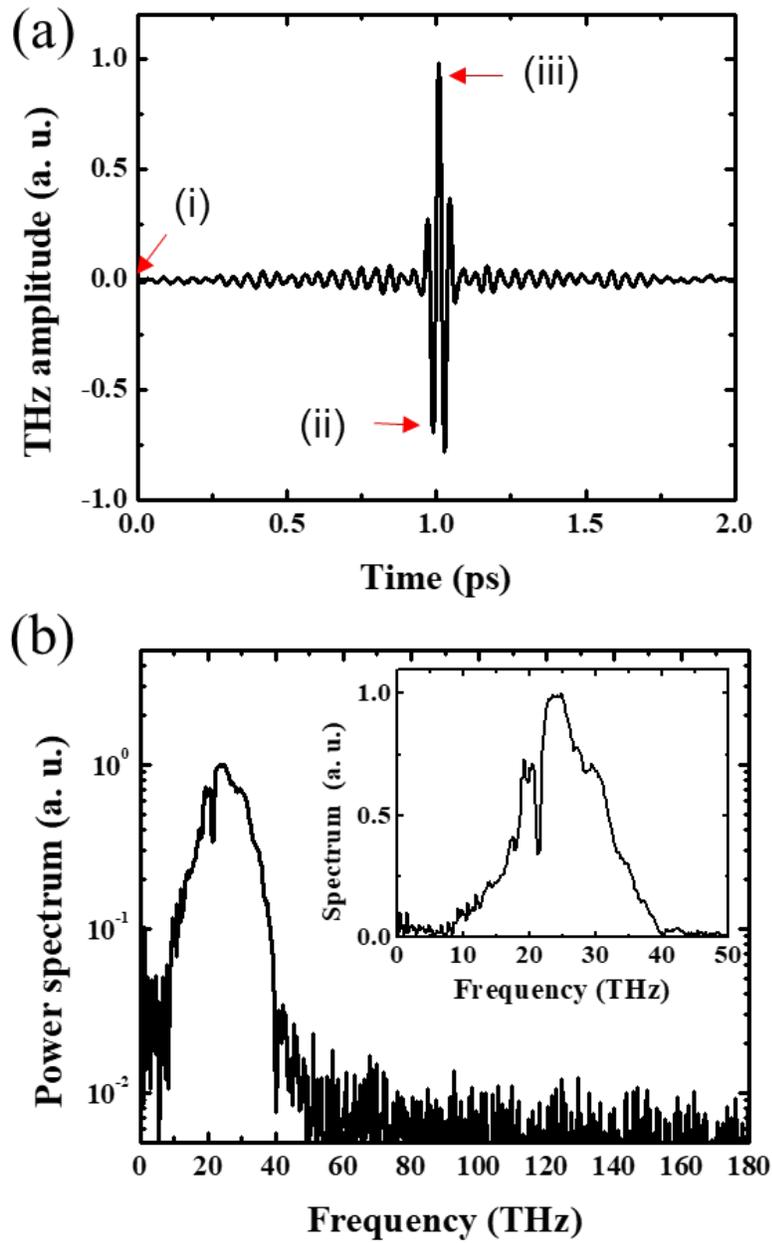

Fig. 5

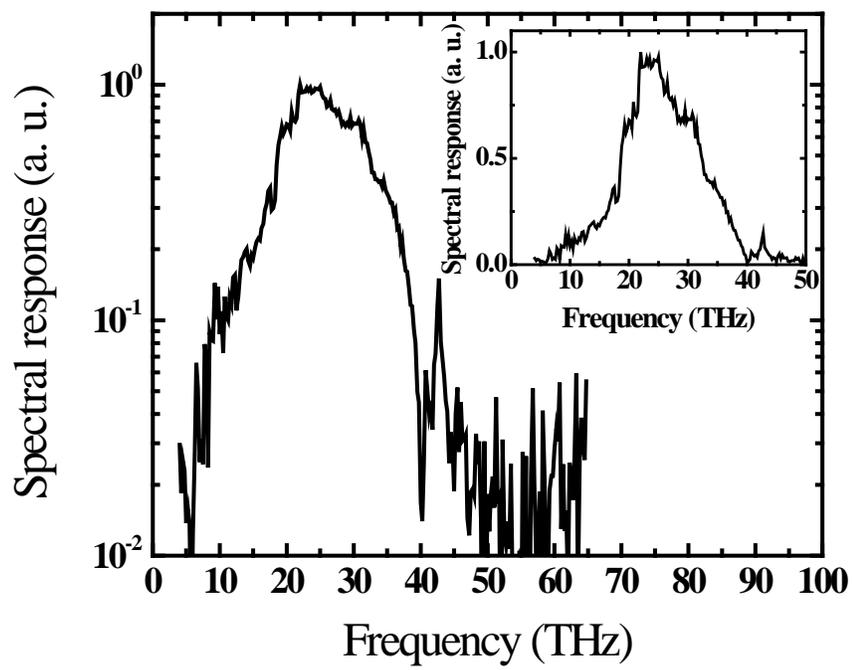

Fig. 6